# Antenna Array Thinning Through Quantum Computing

P. Rocca, [(1)(2)] *Senior Member, IEEE*, N. Anselmi,[(1)] *Member, IEEE*, G. Oliveri,[(1)] *Senior Member, IEEE*, A. Polo,[(1)] *Member, IEEE*, and A. Massa,[(1)(3)(4)] *Fellow, IEEE*

[(1)] *CNIT - "University of Trento" Research Unit*

Via Sommarive 9, 38123 Trento - Italy

E-mail: {*paolo.rocca, nicola.anselmi.1, giacomo.oliveri, alessandro.polo.1, andrea.massa*}@*unitn.it*

Website: *www.eledia.org/eledia-unitn*

[(2)] *ELEDIA Research Center* (*ELEDIA@XIDIAN* - Xidian University)

P.O. Box 191, No.2 South Tabai Road, 710071 Xi'an, Shaanxi Province - China

E-mail: *paolo.rocca@xidian.edu.cn*

Website: *www.eledia.org/eledia-xidian*

[(3)] *ELEDIA Research Center* (*ELEDIA@UESTC* - UESTC)

School of Electronic Engineering, Chengdu 611731 - China

E-mail: *andrea.massa@uestc.edu.cn*

Website: *www.eledia.org/eledia-uestc*

[(4)] *ELEDIA Research Center* (*ELEDIA@TSINGHUA* - Tsinghua University)

30 Shuangqing Rd, 100084 Haidian, Beijing - China

E-mail: *andrea.massa@tsinghua.edu.cn*

Website: *www.eledia.org/eledia-tsinghua*

# Antenna Array Thinning Through Quantum Computing

P. Rocca, N. Anselmi, G. Oliveri, A. Polo, and A. Massa

**Abstract**

Thinning antenna arrays through quantum Fourier transform (*QFT*) is proposed. Given the lattice of the candidate locations for the array elements, the problem of selecting which antenna location has to be either occupied or not by an array element is formulated in the quantum computing (*QC*) framework and then addressed with an *ad-hoc* design method based on a suitable implementation of the *QFT* algorithm. Representative numerical results are presented and discussed to point out the features and the advantages of the proposed *QC*-based thinning technique.

# 1  Introduction

Phased array (*PA*) is a more and more popular antenna technology [1], [2] for a wide range of modern applications such as 5G communications [3]-[5], anti-collision and navigation systems for autonomous driving and unmanned vehicles [6]-[8], future weather and air traffic control radars [9]-[11], and biomedical imaging and therapy [12], [13]. However, minimizing the complexity and the costs of *PA*s is still an open issue of great scientific and industrial interest to enable the diffuse use of *PA*s in a huge number of commercial systems and applications. Towards this end, several studies have been carried out to develop unconventional architectures for cost-effective and high-performance *PA* solutions [14]. Indeed, classical *PA*s consist of bulky and expensive [15] fully-populated arrangements where each antenna element has a dedicated transmit-receive module (*TRM*), composed by an amplifier and a phase shifter for the signal transmission and reception, as well as a radio-frequency (*RF*) chain. Differently, unconventional *PA*s need a reduced number of *TRM*s and *RF* chains since they consider either clustered [16]-[20], or sparse [21]-[24], or thinned [25]-[33] layouts. As for these latter, the radiation pattern is controlled by turning off or removing a set of the elements from an initial fully-populated regular lattice to yield a spatial tapering [1] on the array aperture and, for instance, low sidelobes [2]. The antenna elements are usually uniformly fed so that the *TRM* amplifiers are all saturated to achieve the maximum efficiency and to avoid output power wastening. In this case, only the phase shifters are used for scanning the main beam along the desired direction. Otherwise, alternative thinned architectures exploit tunable amplifiers for controlling the beam shape [34], [35], while off-elements are connected to matched loads instead of being removed. This allows one to maintain a regular distribution of the elements, thus more uniform and predictable electromagnetic coupling conditions.

To select which elements to keep, different design techniques have been proposed. In [25], the thinning of large arrays has been carried out with a random process to match a desired spatial taper, while the arising statistical radiation performance (i.e., sidelobe level, beamwidth, and directivity) have been then studied in [26]. Recently, thanks to the growth of the computational capabilities and the engineering use of advanced stochastic (i.e., global) optimization techniques, thinning methods based on the Simulated Annealing [27], Genetic Algorithms [28], [29], Par-ticle Swarm Optimization [30], and Ant Colony Optimization [31] have been introduced by formulating the synthesis problem as the minimization of a cost function that quantifies the mismatch between the desired radiation performance and the ones iteratively-synthesized. Although global optimization methods theoretically avoid local minima (i.e., sub-optimal solutions) due to the non-convexity of the cost function at hand, their computational cost is usually very high. Dealing with massively thinned arrays [32], analytic methodologies based on Difference Sets (*DS*) and Almost Difference Sets (*ADS*) (i.e., analytic binary sequences that define which elements of the array lattice have to be either off or on) have been also studied since the discrete values of the Fourier transform of the autocorrelation of these binary sequences coincide with samples of the radiation pattern. Such a property has been then exploited to synthesize thinned arrays fitting user-constrained autocorrelation functions instead of power pattern masks [33]. Unfortunately, *DS* and *ADS* sequences are available only for a limited set of array apertures, geometries, and thinning factors[(1)].

This paper presents a novel method for *PA* thinning based on quantum computing (*QC*). Nowadays, *QC* is gaining a growing attention since it proved to be able to drastically speed-up the solution of complex problems when properly formulated within the quantum mechanics framework, which is at the basis of *QC* [36]. Today, the main limitation of *QC* is the realization of quantum computers working with a large number of quantum bits (called *qubits*). Nevertheless, quantum algorithms and approaches have been developed to address, even with classical super-computers, computationally-hard problems in a reasonable amount of time. More specifically, the quantum Fourier transform (*QFT*) [37] is here applied for the first time, to the best of the authors' knowledge, to synthesize thinned antenna arrays. Likewise the discrete Fourier transform (*DFT*), the *QFT* performs a discrete Fourier transform on a list of complex numbers, but the observable (i.e., measurable) outputs are the probabilities of the corresponding quantum states (i.e., real values) [38] instead of the Fourier-transformed coefficients of the input values. However, since the relationship between the coefficients of an array with the elements uniformly-spaced on a rectangular grid and the array factor is a Fourier transform, the idea is to still exploit the *QFT* output to thin the array by keeping the elements of the lattice whose

---

[(1)] The thinning factor is defined as the percentage of active/on elements with respect to all the possible candidate element positions in the lattice.

corresponding quantum state probability is greater than a user-defined threshold. Moreover, the probability of the quantum states is used to define the excitations in case of non-uniform arrays with controllable amplifiers to enable an enhanced control of the arising radiation features.

The rest of the paper is organized as follows. In Sect. 2, the array thinning problem is mathematically formulated in the *QC* framework by presenting the *QFT*-based synthesis method, as well. Selected numerical results are then (Sect. 3) reported to validate the proposed method as well as to analyze and assess its performance in comparison with the classical *DFT* algorithm, as well. Eventually, some conclusions and final remarks are drawn (Sect. 4).

## 2   Mathematical Formulation

Let us consider a linear antenna array of $N$ elements equally-spaced along the $z$-axis, $d$ being the inter-element distance. Each $n$-th ($n = 0, ..., N-1$) element is equipped with a *TRM* that provides an amplitude weight and a phase delay equal to $\alpha_n$ and $\varphi_n$, respectively. The array radiates the following radiation pattern [1]

$$\mathbf{E}(u) = \sum_{n=0}^{N-1} w_n \mathbf{f}_n(u) \exp\left[j(kdu)n\right] \qquad (1)$$

where $w_n$ ($w_n \triangleq \alpha_n \exp[j\varphi_n]$) is the $n$-th ($n = 0, ..., N-1$) complex excitation, $k = \frac{2\pi}{\lambda}$ is the free-space wavenumber, $\lambda$ being the corresponding wavelength, $u$ ($u \triangleq \cos\theta$) is the direction cosine, $\theta$ ($0 \leq \theta \leq \pi$) being the angular direction from the array axis, and $j = \sqrt{-1}$. Under the assumption that the radiation pattern of the $n$-th ($n = 0, ..., N-1$) elementary radiator of the array is equal for all the antennas (i.e., $\mathbf{f}_n(u) = \mathbf{f}(u)$; $n = 0, ..., N-1$), it turns out that $\mathbf{E}(u) \simeq \mathbf{f}(u) \mathcal{A}(u)$, $\mathcal{A}(u)$ being the array factor function given by

$$\mathcal{A}(u) = \sum_{n=0}^{N-1} w_n \exp\left[j(kdu)n\right]. \qquad (2)$$

The thinning of the array is mathematically modeled by introducing a Boolean vector, $\mathbf{B} = \{b_n \in \{0, 1\}; n = 0, ..., N-1\}$. If $b_n = 1$ then the $n$-th ($n = 0, ..., N-1$) *TRM* is connected to the beam-forming network (*BFN*) and the amplitude coefficient is either uniform (i.e., $w_n =$ 1.0) or continuous (i.e., $w_n \in \mathbb{R}_{>0}$, $\mathbb{R}_{>0}$ being the set of positive real numbers greater than zero) for isophoric or non-uniform thinned array, respectively. Otherwise (i.e., $b_n = 0$), the $n$-th ($n = 0, ..., N - 1$) array element has no *TRM* and it is terminated on a matched load (i.e., it does not contribute to the signal transmission and reception through the *BFN*). According to this description, $w_n \to w_n b_n$ ($n = 0, ..., N - 1$) in (2) and the degrees of freedom (*DoF*s) of the thinning problem are the binary entries of the Boolean vector $\mathbf{B}$. The synthesis problem at hand can be then stated as follows:

> *Antenna Array Thinning Problem* (*AATP*) - Given a uniform lattice of $N$ candidate locations equally-space by $d$, determine the $N$ entries of the Boolean vector $\mathbf{B}$ to minimize the cost function $\Psi(\mathbf{B})$ defined as the square distance between a 'feature' of the reference pattern, $\mathbb{G}\left\{\mathcal{A}^{ref}(u)\right\}$, and that of the thinned one, $\mathbb{G}\left\{\mathcal{A}(u \,|\, \mathbf{B})\right\}$
>
> $$\Psi(\mathbf{B}) = \int_{-1}^{1} \left|\mathbb{G}\left\{\mathcal{A}(u \,|\, \mathbf{B})\right\} - \mathbb{G}\left\{\mathcal{A}^{ref}(u)\right\}\right|^2 du. \tag{3}$$

In order to solve the *AATP*, a design strategy based on the *QFT* is presented hereinafter starting from the following observations:

- the relationship between the samples of the array pattern function and the array excitations is a *DFT*;

- the *DFT* of the array excitations provides the least mean squared error approximation of the desired pattern when $d \geq \frac{\lambda}{2}$ [1], [2];

- the *QFT* algorithm is the natural extension to the quantum regime of the *DFT* with a computational complexity of order $\mathcal{O}\left((\log N)^2\right)$ instead of $\mathcal{O}(N \times \log N)$ of the classical *DFT*. It means that the exploitation of the quantum superposition and parallelism can provide an exponential speed-up with respect to standard implementations;

- unlike the *DFT*, the *QFT* gives the probability (i.e., real positive numbers) of the Fourier-transformed input set instead of the complex Fourier values.

Accordingly, the following method has been derived. From (2), it turns out that the sequence/discrete-dataset of $N$ complex samples of the array factor, $\mathbf{A}$ ($\mathbf{A} = \{\mathcal{A}_m; m = 0, ..., N-1\}$), is the *DFT* of the sequence/discrete-dataset of the $N$ complex excitations, $\mathbf{w}$ ($\mathbf{w} = \{w_n; n = 0, ..., N-1\}$),

$$\mathbf{A} = DFT\{\mathbf{w}\} \quad (4)$$

so that the $m$-th ($m = 0, ..., N-1$) entry of $\mathbf{A}$, $\mathcal{A}_m$, ($\mathcal{A}_m \triangleq \mathcal{A}(u_m)$, $u_m$ being the $m$-th sample along the angular cosine direction given by $u_m = -m\frac{\lambda}{N \times d}$ [1]) is equal to [1]

$$\mathcal{A}_m = \sum_{n=0}^{N-1} w_n \exp\left[-j\left(\frac{2\pi}{N}n\right)m\right]. \quad (5)$$

Once the pattern samples are defined, $\mathbf{A}$, the array factor function (2) is yielded as the summation of weighted periodic sinc functions

$$\mathcal{A}(u) = \sum_{m=0}^{N-1} \mathcal{A}_m \mathbb{S}\left(\pi d u + \frac{m\pi}{N}\right) \quad (6)$$

being $\mathbb{S}(x) \triangleq \frac{\sin(Nx)}{N\sin(x)}$.

Vice-versa, the complex excitation set $\mathbf{w}$ is the inverse discrete Fourier transform (*IDFT*) of the discrete dataset of the pattern samples, $\mathbf{A}$

$$\mathbf{w} = IDFT\{\mathbf{A}\} \quad (7)$$

being $w_n = \frac{1}{N}\sum_{m=0}^{N-1} \mathcal{A}_m \exp\left[j\left(\frac{2\pi}{N}m\right)n\right]$ and the array factor function (2) is obtained by substituting $\mathbf{w}$ in (2).

Moving to the quantum framework, let us define the *input quantum state vector* $|\mathbf{A}\rangle$ starting from the $N$-size set of complex pattern samples $\mathbf{A}$ and considering $L$ ($L \triangleq \log_2 N$) single-qubits/1-qubits, $\{|s_\ell\rangle; \ell = 1, ..., L\}$, each having two possible states $|s_\ell\rangle = \{|0\rangle, |1\rangle\}$ (called *kets* in the language of the Dirac notation [38])

$$|\mathbf{A}\rangle = \sum_{m=0}^{N-1} \widehat{\mathcal{A}}_m |\mathbf{a}_m\rangle \quad (8)$$

where $\widehat{\mathcal{A}}_m$ ($\widehat{\mathcal{A}}_m \triangleq \frac{\mathcal{A}_m}{\|\mathbf{A}\|}$, $\|.\|$ being the norm operator) is the complex amplitude of the $m$-th ($m = 0, ..., N - 1$) $L$-qubit, $|\mathbf{a}_m\rangle$, given by $|\mathbf{a}_m\rangle = |s_L \ s_{L-1} \ ... \ s_\ell \ ... \ s_1\rangle$ where $|s_L \ s_{L-1} \ ... \ s_\ell \ ... \ s_1\rangle$ stands for the concatenation of the of the $L$ single-qubits, $\{|s_\ell\rangle; \ \ell = 1, ..., L\}$ ($|s_L s_{L-1} ... s_\ell ... s_1\rangle \triangleq \bigotimes_{\ell=L}^{1} |s_\ell\rangle$)[2]. Analogously to the classical theory, the *output quantum state vector* $|\mathbf{w}\rangle$

$$|\mathbf{w}\rangle = \sum_{n=0}^{N-1} \widehat{w}_n |\mathbf{a}_n\rangle \qquad (9)$$

is the *IQFT* of the *input quantum state vector* $|\mathbf{A}\rangle$ (i.e., $|\mathbf{w}\rangle = IQFT\{|\mathbf{A}\rangle\}$) where $\widehat{w}_n$ ($n = 0, ..., N - 1$) is a normalized complex value equal to $\widehat{w}_n = \frac{w_n}{\|\mathbf{w}\|}$, $w_n$ being derived by $\mathbf{A}$ as in (7) through *IDFT*.

However, it worth pointing out that there is no way to retrieve/measure/observe the (complex) values of the $N$ entries of $\widehat{\mathbf{w}}$ ($\widehat{\mathbf{w}} = \{\widehat{w}_n; n = 0, ..., N - 1\}$) at the output of the *IQFT* operation (i.e., from $|\mathbf{w}\rangle$), but only the probability $p_n$ ($n = 0, ..., N - 1$) that the output quantum state vector $|\mathbf{w}\rangle$ would be equal to the $n$-th ($n = 0, ..., N - 1$) $L$-qubit $|\mathbf{a}_n\rangle$. On the other hand, the following relation between the probability $p_n$ ($n = 0, ..., N - 1$) and $\widehat{w}_n$ ($n = 0, ..., N - 1$) holds true [38]

$$p_n = |\widehat{w}_n|^2. \qquad (10)$$

Therefore, the thinning of the array is performed according to the following multi-step procedure:

- *Step 1 - **Initialization.*** Given the number of candidate array elements $N$ and the inter-element distance $d$, define the reference array pattern, $\mathcal{A}^{ref}(u)$, and choose the desired pattern 'feature', $\mathbb{G}\{\mathcal{A}^{ref}(u)\}$, to be matched in the least-square sense (3) within a user-defined confidence threshold $\eta$ (i.e., $\Psi(\mathbf{B}) \leq \eta$);

- *Step 2 - **QFT Application.*** Apply the *IQFT* to the input quantum state vector $|\mathbf{A}\rangle$ (8) and measure the probability $p_n$ ($n = 0, ..., N - 1$) associated to the output quantum state vector $|\mathbf{w}\rangle$ (9);

- *Step 3 - **Probability Ranking.*** Order the $N$ observed probabilities (10) of the output quan-

---

[2] In an explicit form: $\bigotimes_{\ell=L}^{1} |s_\ell\rangle \triangleq |s_L\rangle \otimes ... \otimes |s_\ell\rangle ... \otimes ... \otimes |s_1\rangle$, $\otimes$ being the tensor product.

tum state vector from the highest $\hat{p}_t]_{t=0} = \max_n \{p_n\}$ down to the lowest $\hat{p}_t]_{t=N-1} = \min_n \{p_n\}$ ($0 \leq t \leq N-1$). In case of equal probability values, rank them according to the value of the $n$-th ($n = 0, ..., N-1$) index, namely $\hat{p}_t = p_{n_1}$ and $\hat{p}_{t+1} = p_{n_2}$ if $p_{n_1} = p_{n_2}$ and $n_1 < n_2$ ($n_1, n_2 \in [0 : N-1]$);

- *Step 4 - **Thinned Array Synthesis.*** Keep 'on' the minimum number of first $K$ elements among the $N$ candidates ($K \leq N$) of the ordered list defined at *Step 3* whose corresponding thinned array (i.e., boolean vector $\mathbf{B}^{opt}$) affords a radiation pattern, $\mathcal{A}(u \,|\, \mathbf{B}^{opt})$, that satisfies the matching condition

$$\Psi\left(\mathbf{B}^{opt}\right) \leq \eta, \tag{11}$$

$\eta$ being the matching threshold, where the $n$-th ($n = 0, ..., N-1$) entry of the thinning Boolean vector $\mathbf{B}^{opt}$ is set to $b_n = 1$ ($b_n = 0$) if the value of the corresponding $t$-th probability ($n \leftarrow t$) lies (does not lie) within the range $0 \leq t \leq K - 1$

$$b_n]_{n \leftarrow t} = \begin{cases} 1 & if \ 0 \leq t \leq (K-1) \\ 0 & if \ K \leq t \leq (N-1) \end{cases}. \tag{12}$$

Moreover, set the $n$-th ($n = 0, ..., N-1$) amplitude excitation to either $w_n = 1.0$ (uniformly-excited array) or to $w_n = \sqrt{p_n}$ (non-uniformly excited array) if $b_n = 1$, while turn off the $n$-th ($n = 0, ..., N-1$) elements otherwise (i.e., $b_n = 0$).

If there is no thinned solution $\mathbf{B}$ that fits (11), relax the value of $\eta$ or, if doable having a wider antenna aperture, increase $N$.

# 3 Numerical Validation

In this section, the *QFT*-based approach for thinning antenna arrays is firstly validated, then selected numerical results are reported and discussed to analyze the robustness of the *QC*-thinning method against the noise on the input data and to assess its performance also in comparison with the classical *DFT*-based implementation. The quantum numerical simulations have been carried out within the Qiskit open source framework [39] by using the IBM Quantum cloud platform [40]. In all simulations, ideal radiating elements (i.e., $\mathbf{f}(u) = 1$) have been considered to focus the attention on the array factor, which is the function optimized by the proposed synthesis method.

In the first example (*Validation*), the linear thinned array has been assumed to be known. It has been generated by thinning an isophoric ($w_n = 1$; $n = 0, ..., N - 1$) fully-populated array lattice with $N = 1024$ equi-spaced ($d = \frac{\lambda}{2}$) locations. More specifically, the $K = 7$ elements with indexes $n = \{0, 1, 3, 4, 5, 7, 9\}$ have been turned on [i.e., $b_n = 1$ - black pixel in Fig. 1(*a*)], while the remaining ones have been turned off [i.e., $b_n = 0$ - white pixel in Fig. 1(*a*)]. Figure 1 shows the corresponding samples of the reference array factor, $\mathbf{A}^{ref} = \{ \mathcal{A}_m^{ref}; m = 0, ..., N - 1\}$, which have been computed through (5) [Fig. 1(*b*)] along with the normalized power pattern values, $\mathbf{P}^{ref} = \{ \mathcal{P}_m^{ref}; m = 0, ..., N - 1\}$ ($\mathcal{P}_m \triangleq |\mathcal{A}_m|^2$) [Fig. 1(*c*)]. The *QC*-thinning method has been then applied to solve the *AATP* where the same array factor function is the pattern 'feature' at hand [i.e., $\mathbb{G}\left\{\mathcal{A}^{ref}(u)\right\} = \mathcal{A}^{ref}(u)$]. Starting from $\mathbf{A}^{ref}$ and applying a cyclic shift such that $m \leftarrow \left(m + \frac{N}{2}\right) \mod N$, the normalized complex amplitudes of the *input quantum state vector* $|\mathbf{A}^{ref}\rangle$, $\widehat{\mathbf{A}}^{ref}$ ($\widehat{\mathbf{A}}^{ref} = \{ \widehat{\mathcal{A}}_m^{ref}; m = 0, ..., N - 1\}$), have been computed and the *IQFT* has run, by using $L = 10$ qubits, $R = 2 \times N$ times to statistically validate the arising thinned array solution. Figure 2 gives the values, measured at the output of the *QC* processor simulated within the Qiskit framework, of the probabilities of the output quantum state vector $|\mathbf{w}\rangle$, $\{p_n; n = 0, ..., N - 1\}$. As it can be observed, only the states corresponding to the indexes of the $K = 7$ elements that were active (i.e., $b_n = 1$) in the reference thinned array have a probability $p_n$ different from zero, while the probabilities of all the others ($N - K$) quantum output states are exactly zero. Moreover, the fact that the probabilities associated to the active elements are equal confirms the effectiveness of the QC-based method in retrieving the isophoric reference arrangement. As for the reliability of the *QC*-based thinning, it is worth pointing out that the same result has been obtained running the *IQFT* for $R = \frac{N}{2}$, $R = N$, and $R = 4 \times N$ times. Of course, a final proof on this matter would be available only running the SW code on a real quantum processor since Qiskit is a realistic emulation tool of a *QC* machine. As for the computational issues, the acceleration factor enabled by the *IQFT* with respect to the use of classical *IDFT* to perform the same four-step thinning process is proportional to the ratio $\left(\frac{N}{\log N}\right)$ and, for this example with $N = 1024$, it amounts to $147$ times.

The second example (*Analysis*) is devoted to check the robustness of the *QC* thinning to the noise blurring the amplitude of the input quantum vector, $\widehat{\mathbf{A}}^{ref}$. Towards this purpose, the $m$-th ($m = 0, ..., N-1$) noisy amplitude $\widehat{\mathcal{A}}_m^{ref}$ has been computed as follows

$$\widetilde{A}_m^{ref} = \widehat{\mathcal{A}}_m^{ref} + \nu_m \tag{13}$$

where $\nu_m$ is the $m$-th realization of an additive zero-mean complex Gaussian noise with variance $\sigma$. By considering different values of the signal-to-noise ratio (*SNR*)

$$SNR \triangleq \frac{\sum_{m=0}^{N-1} \left|\widehat{\mathcal{A}}_m^{ref}\right|^2}{\sum_{m=0}^{N-1} |\nu_m|^2} \tag{14}$$

in the range from $50$ [dB] down to $0$ [dB] with a step of $10$ [dB], each quantum simulation has been repeated $R = 2 \times N$ times. It turns out that the highest probabilities are always associated to the quantum states of the output vector corresponding to the $K = 7$ active elements of the reference thinned array. On the contrary, the probability values of the other $(N - K)$ elements quickly decrease as shown in in the sorted list of Fig. 3. More in detail, they reduce of more than one ($SNR = 0$ [dB]), three ($SNR = 10$ [dB]), four ($SNR = 20$ [dB]), five ($SNR = 30$ [dB]), and six ($SNR \geq 40$ [dB]) orders of magnitude (Fig. 3), thus the corresponding antenna elements are kept off (i.e., $b_n = 0$, $n = 0, ..., N - 1$ and $n \neq \{0, 1, 3, 4, 5, 7, 9\}$) in the final layout of the synthesized thinned array.

In the third example (*Assessment*), the goal of the synthesis is that of thinning a reference lattice of $N = 256$ candidate locations equally-spaced by $d = \frac{\lambda}{2}$ to afford a pattern with a user-defined sidelobe level (*SLL*) (i.e., $\mathbb{G}\left\{\mathcal{A}^{ref}(u)\right\} = SLL_{ref}$) defined as $SLL \triangleq -10 \times \log\left[\max_{u \in \Omega} \left\{\frac{\mathcal{P}^{ref}(0)}{\mathcal{P}^{ref}(u)}\right\}\right]$, $\Omega$ and $\mathcal{P}(u)$ being the sidelobe region and the normalized power pattern ($\mathcal{P}(u) \triangleq |\mathcal{A}(u)|^2$), respectively. Accordingly, $L = 8$ qubits have been used for the *IQFT*, while the non-uniform amplitudes of the array have been set to $\widehat{w}_n = \sqrt{p_n}$ ($n = 0, ..., N - 1$) (10). A wide set of synthesis tests has been carried out by varying the matching threshold $\eta$ (i.e., the degree-of-matching with the requirement) for different values of $SLL_{ref}$. For illustrative purposes, Figure 4 shows some representative samples of the *QC*-synthesized lay-outs, while the corresponding patterns are reported in Fig. 5. Moreover, the values of the descriptive/performance indexes for each thinned solution in Fig. 4 are summarized in Tab. I. As expected, the better the requirement fitting (i.e., a lower value of $\eta$), the greater the number of 'on' elements in the reference lattice, $K$, and the lower the thinning percentage $\tau$ ($\tau \triangleq \frac{N-K}{N}$) (Tab. I). For instance, let us focus on the case of the requirement $SLL_{ref} = -12.5$ [dB]. Decreasing the matching threshold $\eta$ from $\eta = 5.0 \times 10^{-2}$ down to $\eta = 1.25 \times 10^{-2}$, the number of mask violation $\aleph$ (i.e., the number of $u$ values for which $\mathcal{P}(u)\rfloor_{u \in \Omega} > SLL_{th}$) reduces [i.e., $\aleph\rfloor_{\eta=5.0\times 10^{-2}}^{SLL_{ref}=-12.5\,[\text{dB}]} = 38$ - Fig. 5(d), $\aleph\rfloor_{\eta=2.5\times 10^{-2}}^{SLL_{ref}=-12.5\,[\text{dB}]} = 20$ - Fig. 5(e), and $\aleph\rfloor_{\eta=1.25\times 10^{-2}}^{SLL_{ref}=-12.5\,[\text{dB}]} = 12$ - Fig. 5(f)] as well as the average amplitude of the sidelobes $\overline{SLL}$ ($\overline{SLL} \triangleq \frac{1}{2}\int_{u \in \Omega} \mathcal{P}(u)\,du$) [i.e., $\overline{SLL}\rfloor_{\eta=5.0\times 10^{-2}}^{SLL_{ref}=-12.5\,[\text{dB}]} = -16.1$ [dB] - Fig. 5(d), $\overline{SLL}\rfloor_{\eta=2.5\times 10^{-2}}^{SLL_{ref}=-12.5\,[\text{dB}]} = -17.1$ [dB] - Fig. 5(e), and $\overline{SLL}\rfloor_{\eta=1.25\times 10^{-2}}^{SLL_{ref}=-12.5\,[\text{dB}]} = -18.4$ [dB] - Fig. 5(f)], but the thinning percentage decreases (i.e., $\tau\rfloor_{\eta=5.0\times 10^{-2}}^{SLL_{ref}=-12.5\,[\text{dB}]} = 85.16$ %, $\tau\rfloor_{\eta=2.5\times 10^{-2}}^{SLL_{ref}=-12.5\,[\text{dB}]} = 81.64$ %, and $\tau\rfloor_{\eta=1.25\times 10^{-2}}^{SLL_{ref}=-12.5\,[\text{dB}]} = 76.95$ % - Tab. I) since less elements are turned off [Fig. 4(d) vs. Fig. 4(e) vs. Fig. 4(f)]. On the other hand, keeping fixed the $\eta$ value, while decreasing the $SLL$ threshold $SLL_{ref}$, implies that more elements need to be activated to maintain almost constant the number of mask violations $\aleph$ (e.g., $\eta = 2.5 \times 10^{-2}$: $K\rfloor_{SLL_{ref}=-10\,[\text{dB}]}^{\eta=2.5\times 10^{-2}} = 29$ [Fig. 4(b)] $\rightarrow \aleph\rfloor_{SLL_{ref}=-10\,[\text{dB}]}^{\eta=2.5\times 10^{-2}} = 18$, $K\rfloor_{SLL_{ref}=-12.5\,[\text{dB}]}^{\eta=2.5\times 10^{-2}} = 47$ [Fig. 4(e)] $\rightarrow \aleph\rfloor_{SLL_{ref}=-12.5\,[\text{dB}]}^{\eta=2.5\times 10^{-2}} = 20$, and $K\rfloor_{SLL_{ref}=-15\,[\text{dB}]}^{\eta=2.5\times 10^{-2}} = 84$ [Fig. 4(h)] $\rightarrow \aleph\rfloor_{SLL_{ref}=-15\,[\text{dB}]}^{\eta=2.5\times 10^{-2}} = 22$). Finally, the speed-up of the *QC*-based thinning procedure turns out to be 46 times that of the classical one since $N = 256$ for this example.

## 4 Conclusions

An innovative procedure based on QC has been proposed for the synthesis of thinned antenna arrays. Towards this aim, a strategy based on a customization of the *QFT* algorithm has been developed and a set of representative numerical results has been reported and discussed to give some insights on the features and the potentialities of the proposed approach also in comparison with a classical *DFT* implementation.

To the best of the authors' knowledge, the main innovative contributions of this paper with respect to the state-of-the art lie in the following:

- the formulation of the array thinning problem in the *QC* framework;

- the customization of the *QFT* procedure and the exploitation of the observable probabilities of the output quantum state vector for synthesizing both uniformly (i.e., isophoric) and non-uniformly thinned arrays;

- the exploitation of the quantum mechanics principles of superposition and parallelism to yield an exponential acceleration with respect to the classical *DFT* method, thus doing feasible the efficient (i.e., computationally-admissible) thinning of extremely large arrays once quantum processors with many qubits will be available/accessible.

Besides the feasibility of a *QC*-based thinning, the numerical results have assessed the reliability and the effectiveness of the proposed *QFT*-based method in thinning antenna arrays.

As a final remark, it is worth pointing out that the use of *QC*-based procedures for antenna array synthesis and analysis is still an unexplored area of research deserving more studies and investigations. By taking into account that the *DFT* algorithm is at the basis of a wide number of methods for addressing antenna arrays problems [1], [2], it is quite natural that future research activities, beyond the scope of this paper, will be aimed at further leveraging on the peculiarities of *QC* and at exploiting the quantum *supremacy* of its implementations to develop novel breakthrough *QC*-based array methodologies.

# Acknowledgements

This work benefited from the networking activities carried out within the Project "Cloaking Metasurfaces for a New Generation of Intelligent Antenna Systems (MANTLES)" (Grant No. 2017BHFZKH) funded by the Italian Ministry of Education, University, and Research under the PRIN2017 Program (CUP: E64I19000560001). Moreover, it benefited from the networking activities carried out within the Project "SPEED" (Grant No. 61721001) funded by National Science Foundation of China under the Chang-Jiang Visiting Professorship Program, the Project 'Inversion Design Method of Structural Factors of Conformal Load-bearing Antenna Structure based on Desired EM Performance Interval' (Grant no. 2017HZJXSZ) funded by the National

Natural Science Foundation of China, and the Project 'Research on Uncertainty Factors and Propagation Mechanism of Conformal Loab-bearing Antenna Structure' (Grant No. 2021JZD-003) funded by the Department of Science and Technology of Shaanxi Province within the Program Natural Science Basic Research Plan in Shaanxi Province. A. Massa wishes to thank E. Vico for her never-ending inspiration, support, guidance, and help.

# FIGURE CAPTIONS

- **Figure 1.** *Numerical Validation* - Reference thinned layout (*a*). Plot of the $N$ samples of (*b*) the array factor, $\mathbf{A}^{ref} = \{\mathcal{A}_m^{ref}; m = 0, ..., N-1\}$, and of (*c*) the corresponding normalized power pattern, $\mathbf{P}^{ref} = \{\mathcal{P}_m^{ref}; m = 0, ..., N-1\}$ ($\mathcal{P}_m \triangleq |\mathcal{A}_m|^2$).

- **Figure 2.** *Numerical Validation* - Plot of the probability, $p_n$ ($n = 0, ..., N-1$), that the output quantum state vector $|\mathbf{w}\rangle$ is equal to the $n$-th ($n = 0, ..., N-1$) $L$-qubit $|\mathbf{a}_n\rangle$.

- **Figure 3.** *Numerical Analysis* - Plot of the sorted probability value, $p_t$ ($t = 0, ..., N-1$), that the output quantum state vector $|\mathbf{w}\rangle$ is equal to the $t$-th ($t = 0, ..., N-1$) $L$-qubit $|\mathbf{a}_t\rangle$, when blurring the amplitudes of the input quantum vector, $\widehat{\mathbf{A}}^{ref}$, with a Gaussian additive noise characterized by a signal-to-noise ratio equal to $SNR$.

- **Figure 4.** *Numerical Assessment* - *QC*-synthesized thinned layouts when (*a*)(*b*)(*c*) $SLL_{ref} = -10$ [dB], (*d*)(*e*)(*f*) $SLL_{ref} = -12.5$ [dB], and (*g*)(*h*)(*i*) $SLL_{ref} = -15$ [dB] by setting the matching threshold to (*a*)(*d*)(*g*) $\eta = 5.0 \times 10^{-2}$ [dB], (*b*)(*e*)(*h*) $\eta = 2.5 \times 10^{-2}$ [dB], and (*c*)(*f*)(*i*) $\eta = 1.25 \times 10^{-2}$ [dB].

- **Figure 5.** *Numerical Assessment* - Plot of the normalized power patterns radiated by the thinned layouts in Fig. 4.

# TABLE CAPTIONS

- **Table I.** *Numerical Assessment* - Descriptive/performance indexes.

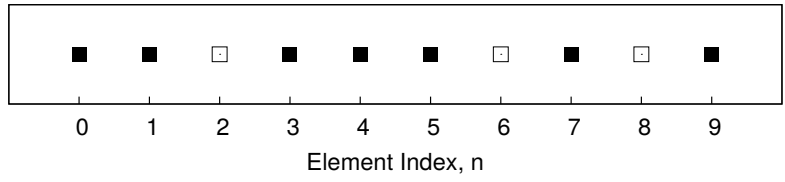

(a)

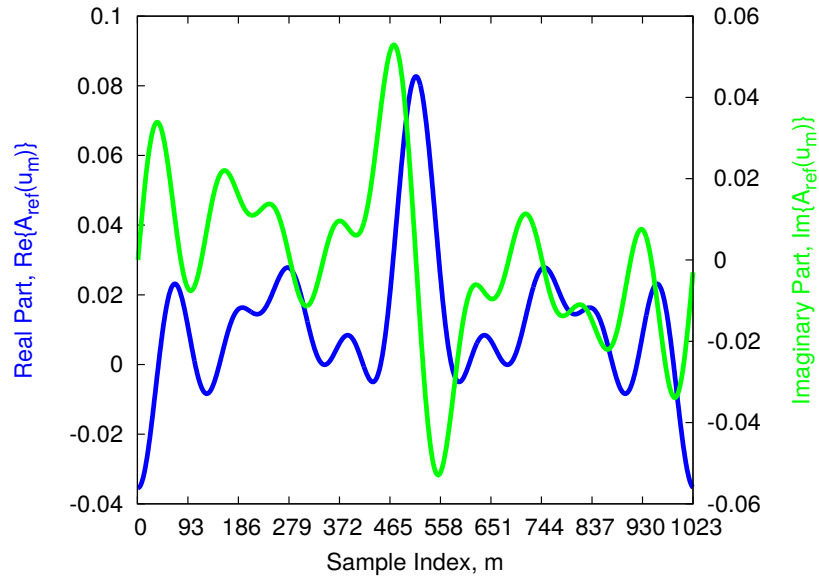

(b)

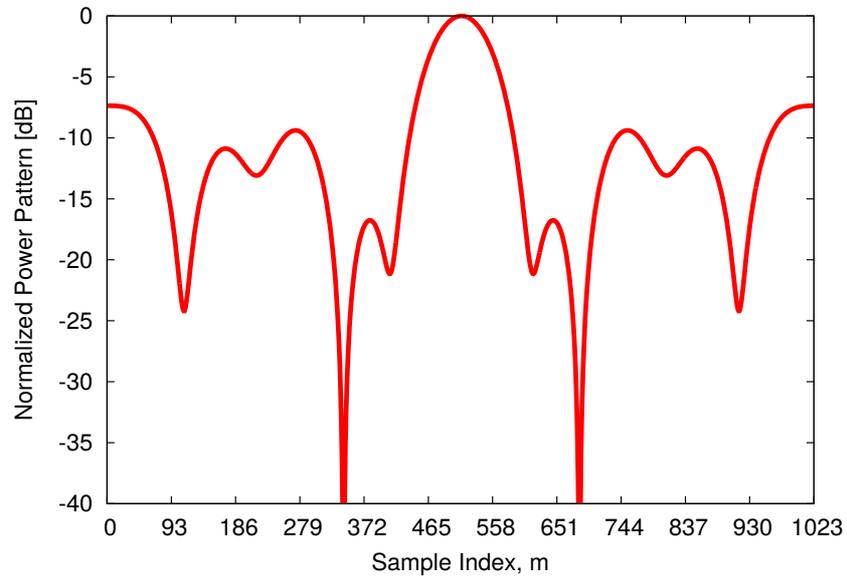

(c)

**Fig. 1** - P. Rocca *et al.*, "Antenna Array Thinning Through Quantum Computing"

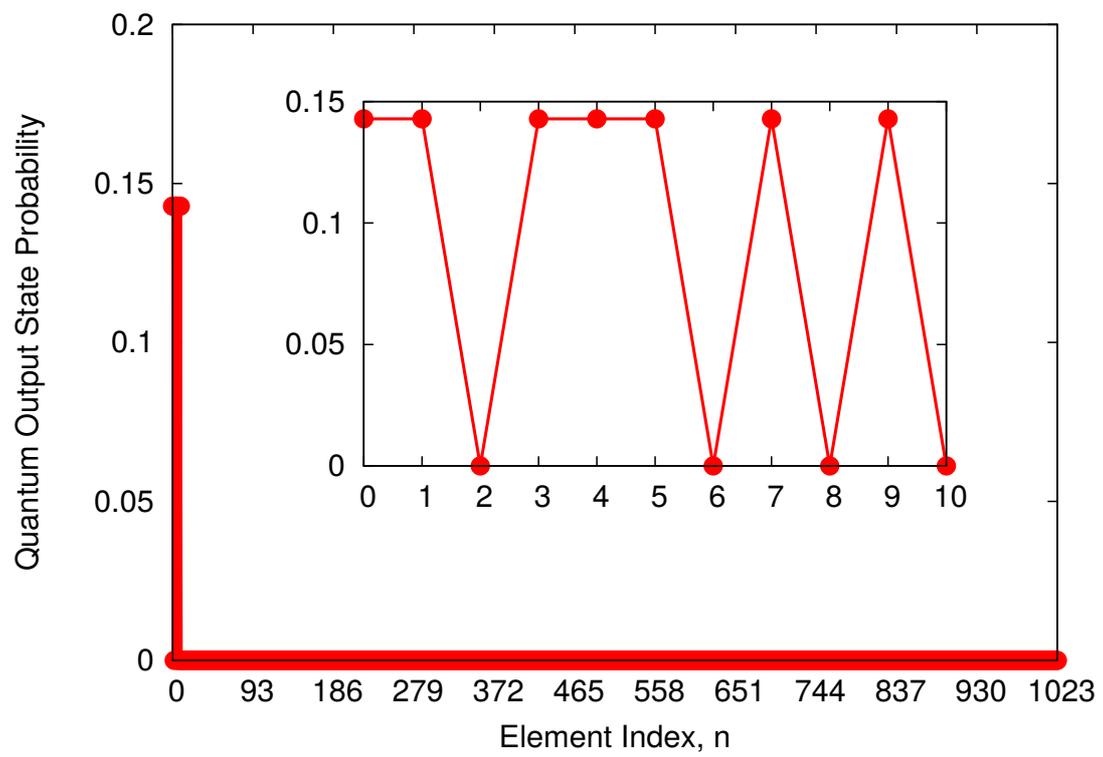

**Fig. 2 -** P. Rocca et *al.*, "Antenna Array Thinning Through Quantum Computing"

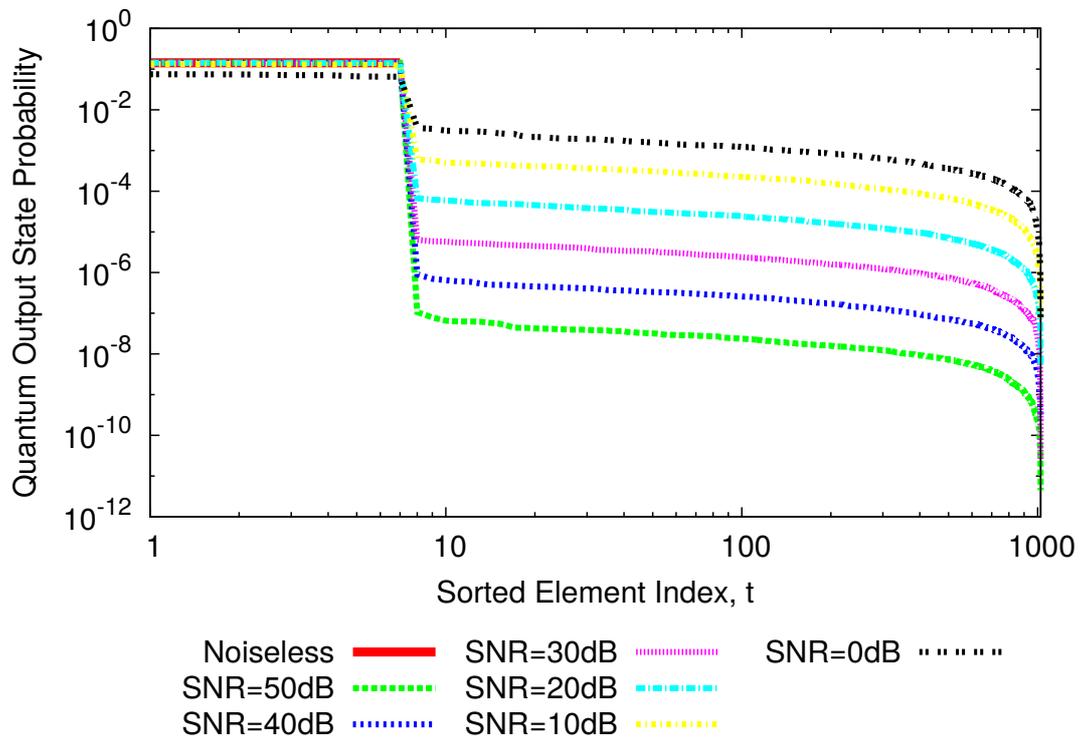

**Fig. 3** - P. Rocca *et al.*, "Antenna Array Thinning Through Quantum Computing"

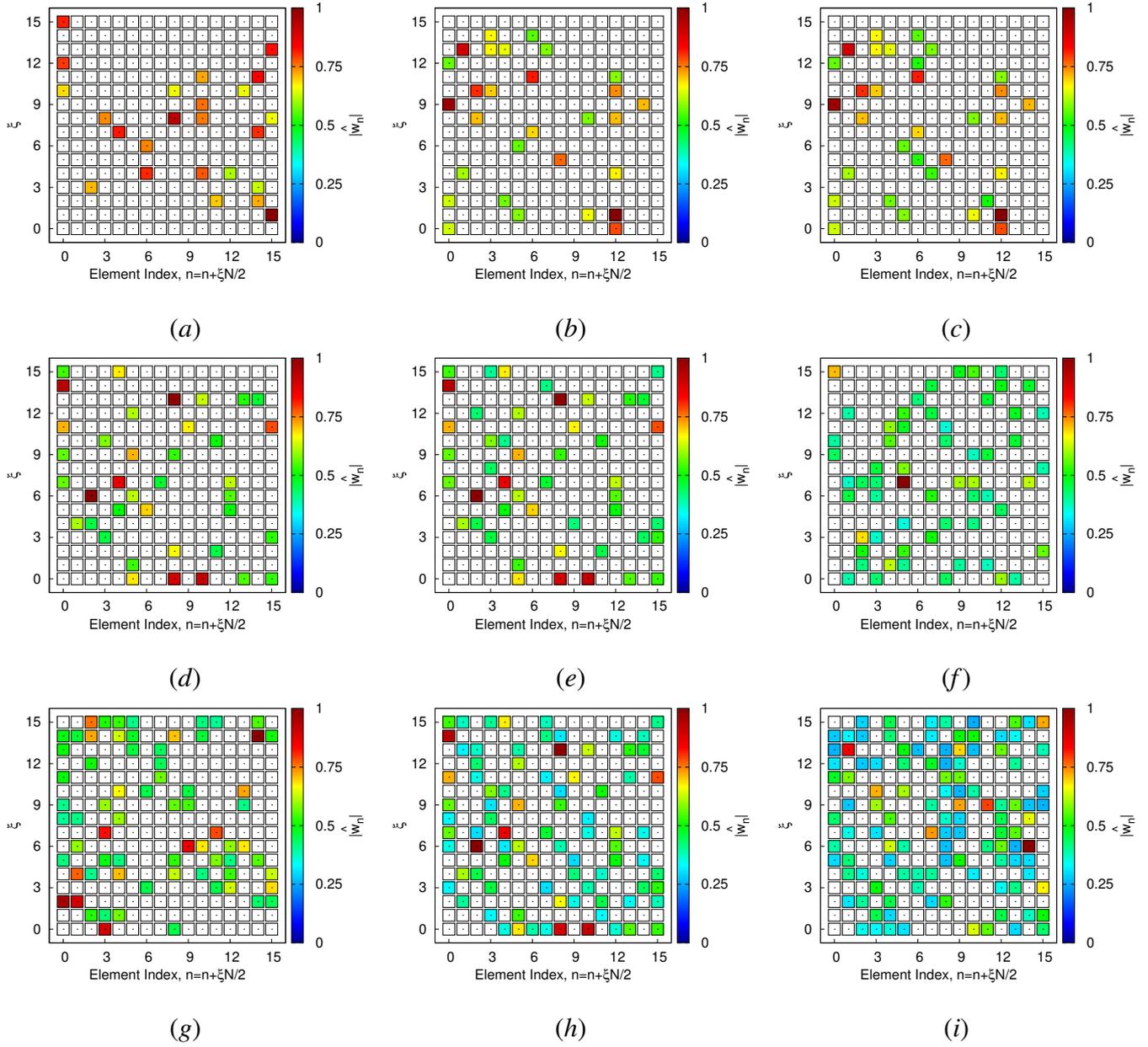

(a) (b) (c)

(d) (e) (f)

(g) (h) (i)

**Fig. 4 - P. Rocca et al.,** "Antenna Array Thinning Through Quantum Computing"

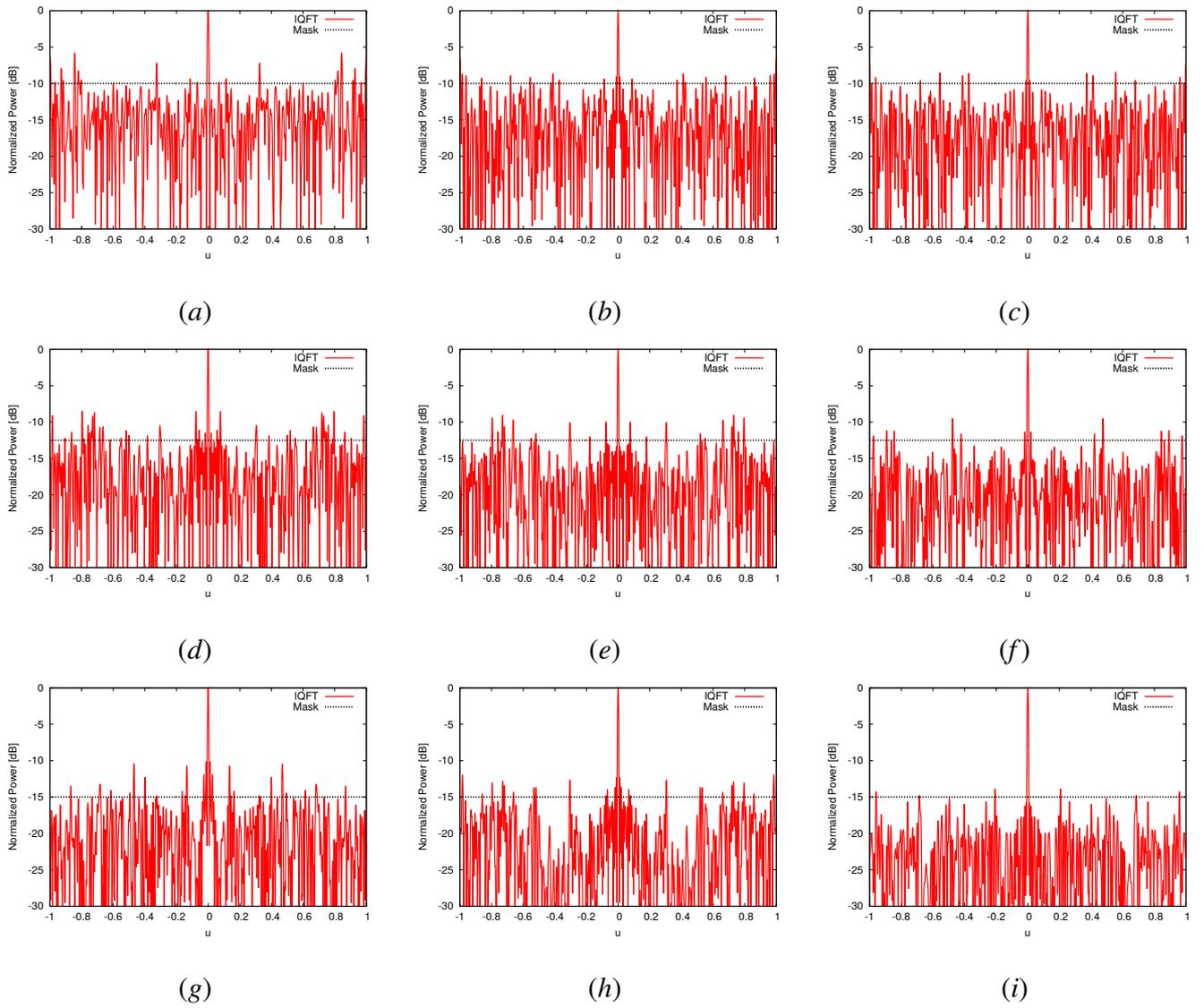

**Fig. 5** - P. Rocca et *al.*, "Antenna Array Thinning Through Quantum Computing"

| $SLL_{ref}$ [dB] | $\eta$ ($\times 10^2$) | $K$ | $\tau$ [%] | $\overline{SLL}$ [dB] | $SLL$ [dB] | $\aleph$ |
|---|---|---|---|---|---|---|
| $-10.0$ | 5.0 | 24 | 90.62 | $-14.2$ | $-5.8$ | 20 |
| $-10.0$ | 2.5 | 29 | 88.67 | $-14.9$ | $-6.5$ | 18 |
| $-10.0$ | 1.25 | 33 | 87.11 | $-15.5$ | $-7.3$ | 12 |
| $-12.5$ | 5.0 | 38 | 85.16 | $-16.1$ | $-8.5$ | 38 |
| $-12.5$ | 2.5 | 47 | 81.64 | $-17.1$ | $-9.1$ | 20 |
| $-12.5$ | 1.25 | 59 | 76.95 | $-18.4$ | $-9.5$ | 12 |
| $-15.0$ | 5.0 | 69 | 73.05 | $-19.2$ | $-10.1$ | 24 |
| $-15.0$ | 2.5 | 84 | 67.19 | $-20.0$ | $-12.0$ | 22 |
| $-15.0$ | 1.25 | 112 | 56.25 | $-21.9$ | $-13.9$ | 6 |

**Table I - P. Rocca et *al.*,** "Antenna Array Thinning Through Quantum Computing"